# Exploiting Commutativity For Practical Fast Replication

Seo Jin Park and John Ousterhout
Stanford University

## Abstract

Traditional approaches to replication require client requests to be ordered before making them durable by copying them to replicas. As a result, clients must wait for two round-trip times (RTTs) before updates complete. In this paper, we show that this entanglement of ordering and durability is unnecessary for strong consistency. Consistent Unordered Replication Protocol (CURP) allows clients to replicate requests that have not yet been ordered, as long as they are commutative. This strategy allows most operations to complete in 1 RTT (the same as an unreplicated system). We implemented CURP in the Redis and RAMCloud storage systems. In RAMCloud, CURP improved write latency by ∼2x (13.8 μs → 7.3 μs) and write throughput by 4x. Compared to unreplicated RAMCloud, CURP's latency overhead for 3-way replication is just 0.4 μs (6.9 μs vs 7.3 μs). CURP transformed a non-durable Redis cache into a consistent and durable storage system with only a small performance overhead.

## 1 Introduction

Fault-tolerant systems rely on replication to mask individual failures. To ensure that an operation is durable, it cannot be considered complete until it has been properly replicated. Replication introduces a significant overhead because it requires round-trip communication to one or more additional servers. Within a datacenter, replication can easily double the latency for operations in comparison to an unreplicated system; in geo-replicated environments the cost of replication can be even greater.

In principle, the cost of replication could be reduced or eliminated if replication could be overlapped with the execution of the operation. In practice, however, this is difficult to do. Executing an operation typically establishes an ordering between that operation and other concurrent operations, and the order must survive crashes if the system is to provide consistent behavior. If replication happens in parallel with execution, different replicas may record different orders for the operations, which can result in inconsistent behavior after crashes. As a result, most systems perform ordering before replication: a client first sends an operation to a server that orders the operation (and usually executes it as well); then that server issues replication requests to other servers, ensuring a consistent ordering among replicas. As a result, the minimum latency for an operation is two round-trip times (RTTs). This problem affects all systems that provide consistency and replication, including both primary-backup approaches and consensus approaches.

Consistent Unordered Replication Protocol (CURP) reduces the overhead for replication by taking advantage of the fact that most operations are commutative, so their order of execution doesn't matter. CURP supplements a system's existing replication mechanism with a lightweight form of replication without ordering based on *witnesses*. A client replicates each operation to one or more witnesses in parallel with sending the request to the primary server; the primary can then execute the operation and return to the client without waiting for normal replication, which happens asynchronously. This allows operations to complete in 1 RTT, as long as all witnessed-but-not-yet-replicated operations are commutative. Non-commutative operations still require 2 RTTs. If the primary crashes, information on witnesses is combined with that from the normal replicas to re-create a consistent server state.

CURP can be easily applied to most existing systems using primary-backup replication. Changes required by CURP are not intrusive, and it works with any kind of backup mechanism (e.g. state machine replication, file writes to network replicated drives, or scattered replication). This is important since most high-performance systems optimize their backup mechanisms, and we don't want to lose those optimizations (e.g. CURP can be used with RAMCloud without sacrificing its fast crash recovery [15]). CURP can even be applied to consensus protocols with a strong leader (such as Raft [14] and Viewstamped Replication [13]).

When CURP is used for geo-replication, it allows consistent update operations in 1 wide-area RTT. At the same time, it also allows strongly consistent reads from local backup replicas (0 wide-area RTTs) through a simple commutativity check with local witnesses.

To show its performance benefits and applicability, we implemented CURP in two NoSQL storage systems: Redis [19] and RAMCloud [16]. Redis is generally used as a non-durable cache due to its very expensive durability mechanism. By applying CURP to Redis, we were able to provide durability and consistency with the similar performance as the current non-durable Redis. For RAMCloud, CURP reduced write latency by half (only a 0.4 μs penalty relative to RAMCloud without replication) and increased throughput by 3.8x without compromising its strong consistency.

Overall, CURP is the first replication protocol that completes linearizable update operations within 1 RTT in asynchronous networks. It can be used for any systems where commutativity of client requests can be checked just from operation parameters (CURP cannot use state-dependent commutativity). Since CURP works correctly in any asynchronous network, it can be deployed to any environment. Even when compared to Speculative Paxos or NOPaxos (which require a special network topology and special network switches), CURP is faster since client request packets do not need to detour to get ordered within networks (NOPaxos has an overhead of 16 µs, but CURP only increased latency by 0.4 µs).

## 2 Separating Durability from Ordering

Replication protocols with concurrent clients have combined the job of ordering client requests consistently among replicas and the job of ensuring the durability of operations. This entanglement causes update operations to take 2 RTTs. CURP achieves consistent ordering and durability separately, so update operations can complete in 1 RTT.

Replication protocols must typically guarantee the following two properties:

- **Consistent Ordering**: if a replica completes operation *a* before *b*, no client in system should see the effects of *b* without that of *a*.
- **Durability**: once its completion has been externalized to an application, an executed operation must survive crashes.

To achieve both consistent ordering and durability, current replication protocols need 2 RTTs. For example, in master-backup (a.k.a. primary-backup) replication, clients' requests are always routed to a master replica, which serializes requests from different clients. As part of executing an operation, the master replicates either the client request itself or the result of the execution to backup replicas; then the master responds back to clients. This entire process takes 2 RTTs total: 1 from clients to masters and another RTT for masters to replicate data to backups in parallel.

Consensus protocols with strong leaders (e.g. Multi-Paxos [7] or Raft [14]) also require 2 RTTs for update operations. Clients route their requests to the current leader replica, which serializes the requests into its operation log. To ensure durability and consistent ordering of the client requests, the leader replicates its operation log to the majority of replicas, and then it executes the operation and replies back to clients with the results. In consequence, consensus protocols with strong leaders also require 2 RTTs for updates: 1 RTT from clients to leaders and another RTT for leaders to replicate the operation log to other replicas.

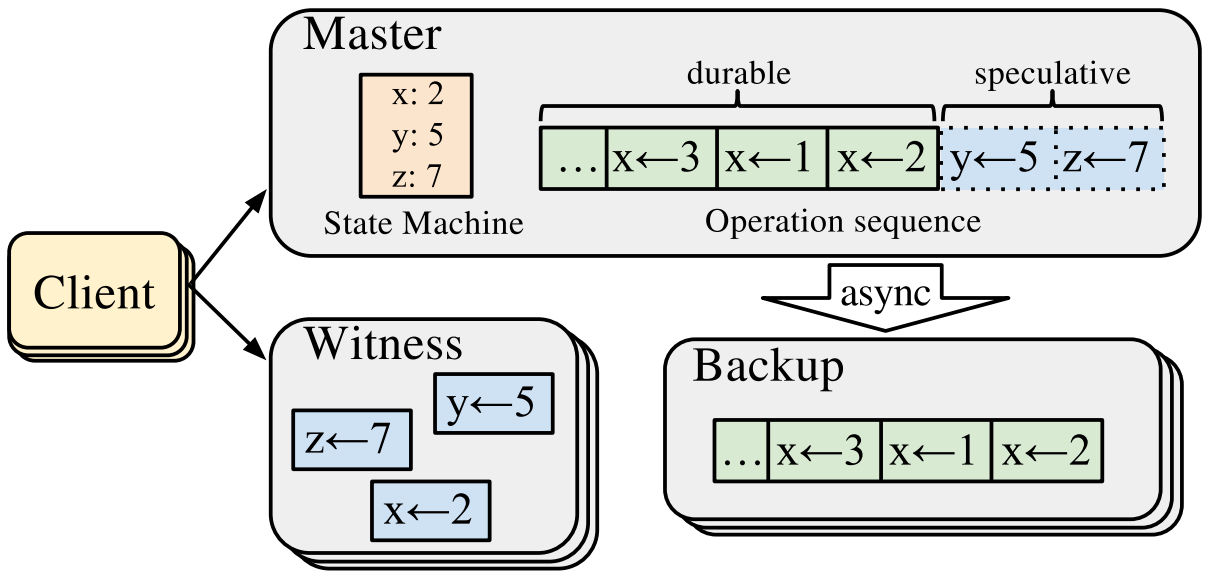

**Figure 1:** CURP clients directly replicate to witnesses. Witnesses only guarantees durability without ordering. Backups hold data that includes ordering information. Witnesses are temporary storage to ensure durability until operations are replicated to backups.

Network-Ordered Paxos (NOPaxos) [11] and Speculative Paxos [17] achieve 1 RTT updates by making special assumptions about the network. They rely on ordered multicast (Ordered Unreliable Multicast and Mostly-Ordered Multicast) to achieve consistent ordering without spending another RTT. However, since both of the protocols require special functions from the network, it is difficult to deploy them in practice. To order them consistently among replicas, client requests in NOPaxos and Speculative Paxos must detour to a common root layer switch. Thus, the protocols are only deployable within a datacenter equipped with a special network topology and advanced switches.

The key idea of CURP is to separate durability and consistent ordering, so update operations can be done in 1 RTT in the normal case. Instead of replicating totally ordered operations in 2 RTTs, CURP achieves durability without ordering and uses the commutativity of operations to defer agreements on operation order.

To achieve durability in 1 RTT, CURP clients directly record their requests in temporary storage, called a *witness*, without serializing them through masters. As shown in Figure 1, witnesses do not carry ordering information, so clients can directly record operations into witnesses concurrently while sending operations to masters (or leaders in consensus). In addition to the unordered replication to witnesses, masters still replicate ordered data to backups, but do so asynchronously after replying the execution results back to clients. Since clients directly make their operations durable through witnesses, masters can reply to clients as soon as they execute the operations without waiting for permanent replication to backups. If a master crashes, the client requests recorded in witnesses are replayed to recover the last few operations that were not replicated to backups. A client can then complete an update operation and reveal the result returned from the master if it successfully recorded the request in witnesses (optimistic fast path: 1 RTT), or after waiting for the master to replicate to backups (slow path: 2 RTT).

CURP's approach introduces two threats to consistency: ordering and duplication. The first problem is that the order in which requests are replayed after a server crash may not match the order in which the master processed those requests. CURP uses commutativity to solve this problem: all of the *speculative* requests (those that a client considers complete, but which have not been replicated to backups) must be commutative. Given this restriction, the order of replay will have no visible impact on system behavior. Specifically, a witness only accepts and saves an operation if it is commutative with every other operation currently stored by that witness (e.g., writes to different objects). In addition, a master will only execute client operations speculatively (by responding before replication is complete), if that operation is commutative with every other speculative operation. If either a witness or master finds that a new operation is not commutative, the client must ask the master to sync with backups. This adds an extra RTT of latency, but it flushes all of the speculative operations.

The second problem introduced by CURP is duplication. When a master crashes, it may have completed the replication of one or more operations that are recorded by witnesses. Any completed operations will be re-executed during replay from witnesses. Thus there must be a mechanism to detect and filter out these re-executions. The problem of re-executions is not unique to CURP, and it can happen in distributed systems for a variety of other reasons. There exist mechanisms to filter out duplicate executions, such as RIFL [10], and they can be applied to CURP as well.

We can apply the idea of separating ordering and durability to both consensus-based replicated state machines (RSM) and master-backup, but this paper focuses on master-backup since it is more critical for application performance. Fault-tolerant large-scale high-performance systems are mostly configured with a single cluster coordinator replicated by consensus and many data servers using master-backup (e.g. Chubby [1], ZooKeeper [6], Raft [14] are used for cluster coordinators in GFS [4], HDFS [20], and RAMCloud [16]). Operations to the cluster coordinators are usually for configuration management (e.g. adding or dropping a table), and they are infrequent and less latency sensitive. On the other hand, operations to data servers (e.g. insert, replace, etc) directly impact application performance, so the rest of this paper will focus on the CURP protocol for master-backup, which is the main replication technique for data servers. In §A.2, we sketch how the same technique can be applied for consensus.

# 3 CURP Protocol

CURP is a new replication protocol that allows clients to complete linearizable updates within 1 RTT. Masters in CURP speculatively execute and respond to clients before the replication to backups has completed. To ensure the durability of the speculatively completed updates, clients multicast update operations to witnesses. To preserve linearizability, witnesses and masters enforce commutativity among operations that are not fully replicated to backups.

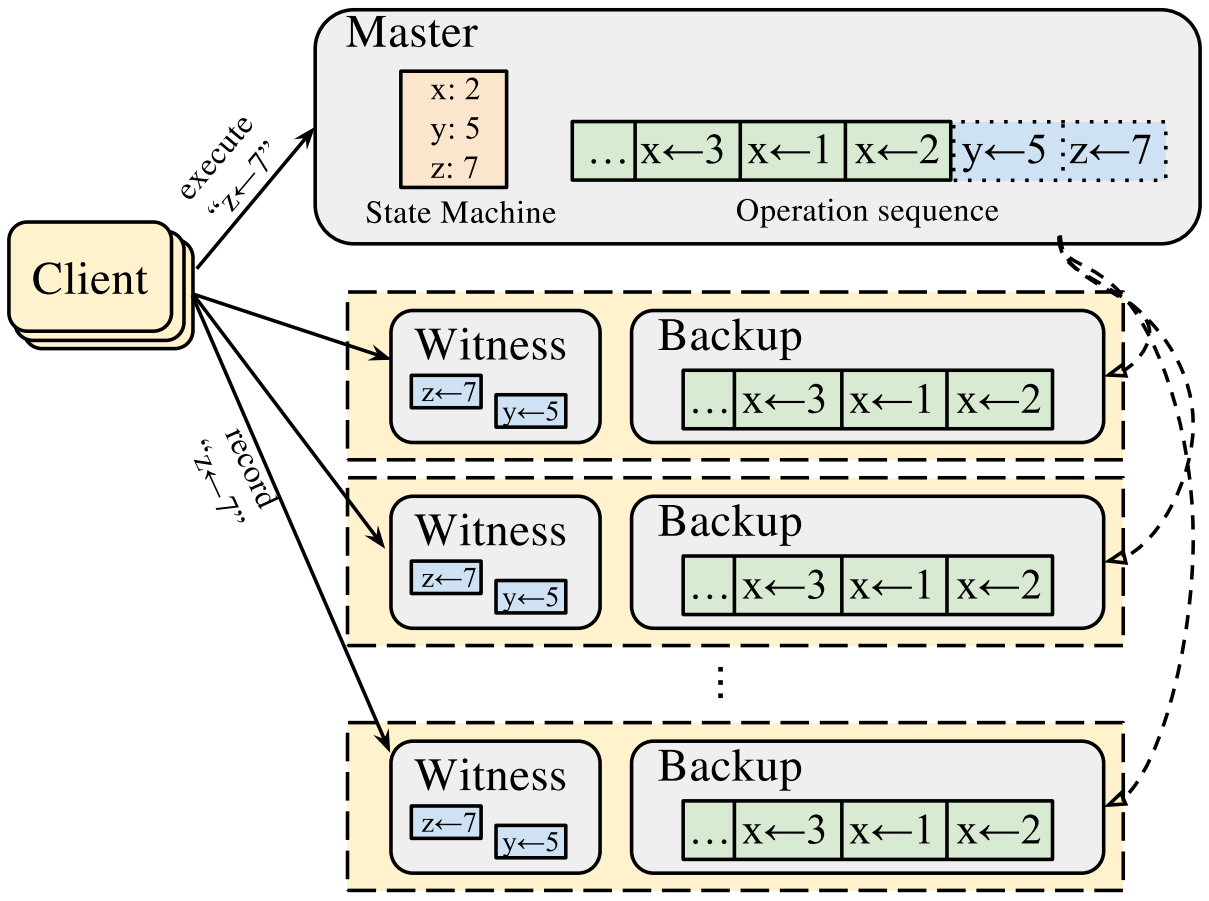

**Figure 2:** CURP architecture

## 3.1 Architecture and Model

CURP provides the same guarantee as current primary-backup protocols; it provides linearizability to client requests in spite of failures. CURP assumes a fail stop model and does not handle byzantine faults. As in typical primary-backup replications, it uses total of $f+1$ replicas composed of 1 master and $f$ backups, where $f$ is the number of replicas that are allowed to fail. In addition to that, it uses $f$ witnesses to ensure durability of updates even before replications to backups are completed (as shown in Figure 2, witnesses are lightweight and can be co-hosted with backups). CURP remains available (i.e. immediately recoverable) with $f$ failures, but will be still strongly consistent even if all replicas fail.

CURP makes no assumptions about the network. It operates correctly even with networks that are asynchronous (no bound on message delay) and unreliable (messages can be dropped) networks. Thus, it can achieve 1 RTT updates on replicated systems in any environment, unlike other alternative solutions. (For example, Speculative Paxos [17] and Network-Ordered Paxos [11] made assumptions on network and cannot be used for geo-replications.)

## 3.2 Normal Operation

This section describes how each component in the system maintains consistency during normal operations.

### 3.2.1 Client

Clients interact with masters generally same as they would without CURP. Clients send update RPC requests to masters. If a client cannot receive a response, it retries the update RPC. If the master crashes, the client may retry the RPC to a different server which recovered the crashed master.

For 1 RTT updates, masters return to clients before replication to backups. To ensure durability, clients directly record their requests to *witnesses* concurrently while waiting for responses from masters. Once all $f$ witnesses have accepted the requests, clients are assured that they will survive master crashes, so clients complete the operations with the results returned from masters.

If a client cannot record in all $f$ witnesses (due to failures or rejections by witnesses), the client cannot complete an update operation in 1 RTT. To ensure the durability of the operation, the client must wait for replication to backups by sending **sync** RPCs to the master. When they receive sync RPCs, masters replicate their current state to backups and then return to clients. If there is no response to the sync RPC (indicating the master might have crashed), the client restarts the entire process; it resends the update RPC to a new master and tries to record the RPC request in witnesses of the new master.

### 3.2.2 Witness

Witnesses support 3 basic operations: they record operations in response to client requests, hold the operations until explicitly told to drop by masters, and provide the saved operations during recovery.

Once a witness accepts the **record** RPC for an operation, it guarantees the durability of the operation until told that the operation is safe to drop. To be safe from power failures, witnesses store their data in non-volatile memory (such as flash-backed DRAM). This is feasible since a witness needs only a small amount of space to temporarily hold recent client requests. Similar techniques are used in strongly-consistent low-latency storage systems, such as RAMCloud [16].

A witness accepts a new **record** RPC from a client only if the new operation is commutative with all operations that are currently saved in the witness. If the new request doesn't commute with one of the existing requests, the witness must reject the record RPC since the witness has no way to order the two noncommutative operations consistent with the execution order in masters. For example, if a witness already accepted "$x \leftarrow 1$", it cannot accept "$x \leftarrow 5$".

Witnesses must be able to determine whether operations are commutative or not just from the operation parameters. For example, in key-value stores, witnesses can exploit the fact that operations on different keys are commutative. In some cases, it is difficult to determine whether two operations commute each other. SQL UPDATE is an example; it is impossible to determine the commutativity of "`UPDATE T SET rate = 40 WHERE level = 3`" and "`UPDATE T SET rate = rate + 10 WHERE dept = SDE`" just from the requests themselves. To determine the commutativity of the two updates, we must run them with real data. Thus, witnesses cannot be used for operations whose commutativity depends on the system state.

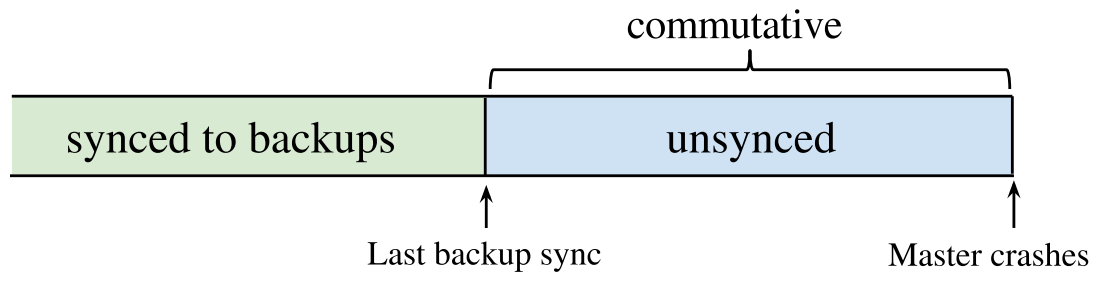

**Figure 3:** Sequence of executed operations in the crashed master.

Each of $f$ witnesses operates independently; witnesses need not agree on either ordering or durability of operations. In an asynchronous network, record RPCs may arrive at witnesses in different order, which can cause witnesses to accept and reject different sets of operations. However, this does not endanger consistency. First, as mentioned in §3.2.1, a client can proceed without waiting for sync to backups only if all $f$ witnesses accepted its record RPCs. Second, requests in each witness are required to be commutative independently, and only one witness is selected and used during the recovery (described in §3.3).

### 3.2.3 Master

The role of masters in CURP is similar to their role in traditional primary-backup replications. Masters in CURP receive, serialize, and execute all update RPC requests from clients. If the executed operation updated the system state, the master synchronizes (can be abbreviated as *sync*) its current state with backups by replicating the updated value or the log of ordered operations.

Unlike traditional primary-backup replications, masters in CURP generally respond back to clients *before* syncing to backups, so that clients can receive the results of update RPCs within 1 RTT. We call this *speculative* execution since the executions may be lost if masters crash. Also, we call the operations that were speculatively executed but not yet replicated to backups *unsynced* operations. As shown in Figure 3, all unsynced operations are contiguous at the tail of the masters' execution history.

To prevent inconsistency, a master must sync before responding if the operation is not commutative with any existing unsynced operations. If a master responds for a noncommutative operation before syncing, the result returned to the client may become inconsistent if the master crashes. This is because the later operation might complete and its result could be externalized (because it was recorded to witnesses) while the earlier operation might not survive the crash (because, for example, its client crashed before recording it to witnesses). For example, if a master speculatively executes "$x \leftarrow 2$" and "read $x$", the returned read value, 2, will not be valid if the master crashes and loses "$x \leftarrow 2$". To prevent such unsafe dependencies, masters enforce commutativity among unsynced operations; this ensures that all results returned to clients will be valid as long as they are recorded in witnesses.

If an operation is synced because of a conflict, the master tags its result as "synced" in the response; so, even if the witnesses rejected the operation, the client doesn't need to send a **sync** RPC. Thus, in most conflict cases, operations complete in 2 RTTs.

### 3.3 Recovery

CURP recovers from a master's crash in two steps: restoration from backups and replay from witnesses. First, the new master restores data from one of the backups, using the same mechanism it would have used in the absence of CURP.

Once all data from backups have been restored, the new master replays the requests recorded in witnesses. The new master picks any available witness. If none of the $f$ witnesses are reachable, the new master must wait. After picking the witness to recover from, the new master first asks it to stop accepting more operations; this prevents clients from erroneously completing update operations after recording them in a stale witness whose requests will not be retried anymore. After making the selected witness immutable, the new master retrieves the requests recorded in the witness. Since all requests in a single witness are guaranteed to be commutative, the new master can execute them in any order. After replaying all requests recorded in the selected witness, the new master finalizes the recovery by syncing to backups and resetting witnesses for the new master (or assigning a new set of witnesses). Then the new master can start to take client requests again.

Some of the requests in the selected witness may have been executed and replicated to backups before the master crashed, so the replay of such requests will result in re-execution of already executed operations. Duplicate executions of the requests can violate linearizability [10].

To avoid duplicate executions of the requests that are already replicated to backups, CURP relies on exactly-once semantics provided by RIFL [10], which detects already executed client requests and avoids their re-execution. Such mechanisms for exactly-once semantics are already necessary to achieve linearizability for distributed systems [10], so CURP is not creating a new requirement. In RIFL, clients assign a unique ID to each RPC; servers save the IDs and results of completed requests and use them to detect and answer duplicate requests. The IDs and results are durably preserved with updated objects in an atomic fashion. (If a system replicates client requests to backups instead of just updated values, providing atomic durability becomes trivial since each request already contains its ID and its result can be obtained from its replay during recovery.)

### 3.4 Safety

With the normal operation behaviors described in §3.2, the recovery protocol in §3.3 guarantees that all client requests are linearizable even after a master crashes.

We first demonstrate that an operation is durable (immediately recoverable with up to $f$ failures) if a client completes it. A client only completes an update operation if (1) it is recorded in all $f$ witnesses or (2) it is replicated to $f$ backups by asking and waiting for the master to sync. Since recovery of a master only completes after recovery from 1 backup and 1 witness, the completed operation must be recovered as long as the recovery is completed.

Secondly, we show that all client requests are linearizable. Linearizability [5] is the strongest form of consistency with concurrent clients, where each operation should *appear* to happen exactly-once at a time point (called the linearization point) between when a client begins the operation and the client ends the operation. If the client crashes before externalizing the result, the RPC may or may not finish.

The commutativity of unsynced operations comes in handy to prove linearizability. We can think of the sequence of operations executed in a master in two parts: operations synced to backups and unsynced operations (see Figure 3). After a master recovery, all unsynced operations that were completed by clients (in other words, recorded in all witnesses) will be recovered. Since all requests that were retained in the witness used for recovery must be commutative, all completed operations are recovered with the exact same outcome.

Some unsynced operations may not have been recorded in the selected witness if the witness rejected them; then the unsynced operations are lost after the master crashes. In this situation, the client could not have completed the operation without waiting for a sync; but the client's sync RPC must have been failed (if they were succeeded, they are not *unsynced* operations), so the client will restart the operation again. On the other hand, if the client crashed, it is safe to lose the operation since the operation was not completed before the crash of the client.

Also, some requests may have been recorded in the selected witness before the crashed master speculatively executed it. During recovery, such requests will be executed and synced to backups. Since the old master crashed before executing the requests, clients retry them to the new master. This does not cause linearizability failure since the retries from clients will not be re-executed (thanks to RIFL [10]), and the saved results from the original executions will be returned to the clients.

### 3.5 Garbage Collection

To limit memory usage in witnesses and reduce possible rejections due to commutativity violations, witnesses must discard requests as soon as possible. Witnesses can drop the recorded client requests after masters make their outcomes durable in backups. In CURP, masters send garbage collection RPCs for the synced updates to their witnesses. The garbage collection RPCs are batched: each RPC lists several operations that are now durable (using RPC IDs provided by RIFL [10]).

### 3.6 Reconfigurations

This section discusses three cases of reconfigurations: recovery of a crashed backup, recovery of a crashed witness, and data migration for load balancing. First, CURP doesn't change the way to handle backup failures, so a system can just recover a failed backup as if it would with the standard primary-backup protocol before adopting CURP.

Second, if a witness crashes or disconnects, the system configuration manager (the owner of all cluster configurations) decommissions the crashed witness and assigns a new witness for the master; then it notifies the master of the new witness list. When the master receives the notification, it syncs to backups to ensure $f$ fault tolerance and responds back to the configuration manager that it is now safe to recover from the new witness. After this point, clients can use $f$ witnesses again to record operations. However, CURP does not push the new list of witnesses to clients. Since clients cache the list of witnesses, clients may still use the decommissioned witness (if it was temporarily disconnected, the witness will continue to accept record RPCs from clients). This endangers consistency since the requests recorded in the old witnesses will not be replayed during recovery.

To prevent clients from completing an unsynced update operation with just recording to old witnesses, CURP maintains a monotonically increasing integer, *WitnessListVersion*, for each master. A master's *WitnessListVersion* is incremented every time when the witness configuration for the master is updated, and the master is notified of the new version along with the new witness list. Clients obtain the *WitnessListVersion* when they fetch the witness list from configuration manager. On all update requests, clients include *WitnessListVersion*s, so that masters can detect and return errors if the clients used wrong witnesses; if they receive errors, the clients fetch new witness lists and retry the updates. Effectively, clients' update operations can never complete without syncing to backups or recording to current witnesses.

Third, for load balancing, a master can split its data into two partitions and migrate a partition to a different master. Migrations usually happen in two steps: a prepare step of copying data while servicing requests and a final step which stops servicing (to ensure that all recent operations are copied) and changes configuration. To simplify the protocol changes from the base primary-backup protocol, CURP masters sync to backups and reset witnesses before the final step of migration, so witnesses are completely ruled out of migration protocols. After the migration is completed, some clients may send updates on the migrated partition to the old master and old witnesses; the old master will reject and tell the client to fetch the new master information (this is the same as before applying CURP); then the client will fetch the new master and its witness information and retry the update. Meanwhile, the requests on the migrated partition can be accidentally recorded in the old witness, but this does not cause safety issues; masters will ignore such requests during replay phase of recovery by the filtering mechanism used to reject requests on not owned partitions during normal operations.

## 4 Implementation on NoSQL Storage

This section describes how to implement CURP on low-latency NoSQL storage systems that use primary-backup replications. With the emergence of large-scale Web services, NoSQL storage systems became very popular (e.g. Redis [19], RAMCloud [16], DynamoDB [21] and MongoDB [2]), and they range from simple key-value stores to more fully featured stores supporting secondary indexing and multi-object transactions; so, improving their performance using CURP is an important problem with a broad impact.

The most important piece missing from §3 to implement CURP is how to efficiently detect commutativity violations. Fortunately for NoSQL systems, CURP can use *primary keys* to efficiently check the commutativity of operations. NoSQL systems store data as a collection of objects, which are identified by *primary key*s. Most update operations in NoSQL specify the affected object with its primary key (or a list of primary keys), and the update operations are commutative if they are modifying disjoint sets of objects. The rest of this section describes an implementation of CURP that exploits this efficient commutativity check.

### 4.1 Life of A Witness

Witnesses have two modes of operation: normal and recovery. In each mode, witnesses service a subset of operations listed in Figure 4. When it receives a **start** RPC, a witness starts its life for a master in a normal mode, in which it is allowed to mutate its collection of saved requests. In the normal mode, the witness services **record** RPCs for client requests targeted to the master for which the witness was configured by **start**; by accepting only requests for the correct master, CURP prevents clients from recording to incorrect witnesses. Also, witnesses drop their saved client requests as they receive **gc** RPCs from masters.

A witness irreversibly switches to a recovery mode once it receives a **getRecoveryData** RPC from a master that is recovering a crashed master. In recovery mode, mutations on the saved request are prohibited; witnesses reject all **record** RPCs and only service **getRecoveryData** or **end**. As a recovery is completed and the witness becomes useless, the cluster coordinator may send **end** to free up the resources, so that the witness server can start another life for a different master.

CLIENT TO WITNESS:

**record**(*masterID*, list of *keyHash*, *rpcId*, *request*) → {ACCEPTED or REJECTED}

Saves the client *request* (with *rpcId*) of an update on *keyHash*es. Returns whether the witness could accomodate and save the request.

MASTER TO WITNESS:

**gc**(list of {*keyHash, rpcId*}) → list of *request*

Drops the saved requests with the given *keyHash*es and *rpcId*s. Returns stale requests that haven't been garbage collected for a long time.

**getRecoveryData**() → list of *request*

Returns all requests saved for a particular crashed master.

CLUSTER COORDINATOR TO WITNESS:

**start**(*masterId*) → {SUCCESS or FAIL}

Start a witness instance for the given master, and return SUCCESS. If the server fails to create the instance, FAIL is returned.

**end**() → *NULL*

This witness is decommissioned. Destruct itself.

**Figure 4:** The APIs of Witnesses.

### 4.2 Data Structure of Witnesses

Witnesses are designed to minimize the CPU cycles spent for handling **record** RPCs. For client requests mutating a single object, recording to a witness is similar to inserting in a set-associative cache; a record operation finds a set of slots using a hash of the object's primary key and writes the given request to an available slot in the set. To enforce commutativity, the witness searches the occupied slots in the set and rejects if there is another request with the same primary key (for performance, we compare 64-bit hashes of primary keys instead of full keys). If there is no slot available in the set for the key, the record operation is rejected as well.

For client requests mutating multiple objects, witnesses perform the commutativity and space check for every affected object; to accept an update affecting *n* objects, a witness must ensure that (1) no existing client request mutates any of the *n* objects and (2) there is an available slot in each set for all *n* objects. If the update is commutative and space is available, the witness writes the update request *n* times as if recording *n* different requests on each object.

### 4.3 Commutativity Checks in Masters

Every NoSQL update operation changes the values of one or more objects. To enforce commutativity, a master can check if the objects touched (either updated or just read) by an operation are *unsynced* at the time of its execution. If an operation touches any *unsynced* value, it is not commutative and the master must sync all unsynced operations to backups before responding back to the client.

If the object values are as stored in a log structure, masters can determine if an object value is synced or not by comparing its position in the log against the last synced position.

If the object values are not stored in a log, masters can use monotonically increasing timestamps. Whenever a master updates the value of an object, it tags the new value with a current timestamp. Also, the master keeps the timestamp of when last backup sync started. By comparing the timestamp of last update of objects against the timestamp of last backup sync, masters can tell whether the value of an object has been synced to backups.

### 4.4 Improving Throughput of Masters

Masters in primary-backup replication are usually the bottlenecks of systems since they drive replication to backups. Since masters in CURP can respond to clients before syncing to backups, they can delay syncs until the next batch without impacting latency. This batching of syncs improves masters' throughput in two ways.

First, by batching replication RPCs, CURP reduces the number of RPCs a master must handle per client request. With 3-way primary-backup replication, a master must process 4 RPCs per client request (1 update RPC and 3 replication RPCs). If the master batches replication and syncs every 10 client requests, it handles 1.3 RPCs on average. On NoSQL storage systems, sending and receiving RPCs takes a significant portion of the total processing time since NoSQL operations are not compute-heavy.

Second, CURP eliminates wasted resources and other inefficiencies that arise when masters wait for syncs. For example, in the RAMCloud [16] storage system, request handlers use a polling loop to wait for completion of backup syncs. The syncs complete too quickly to context-switch to a different activity, but the polling still wastes more than half of the CPU cycles of the polling thread. With CURP, a master can complete a request without waiting for syncing and move on to the next request immediately, which results in higher throughput.

The batch size of syncs is limited in CURP to reduce witness rejections. Delaying syncs increases the chance of finding non-commutative operations in witnesses and masters, causing extra rejections in witnesses and more blocking syncs in masters. To keep the probability of noncommutative operations low, masters batch at most 50 operations before syncs (the number 50 was empirically obtained; larger batches marginally help throughput). In addition, masters sync preemptively after executing an update on a object that had been updated recently as well (this hints it will be updated again soon); this heuristic prevents future requests on the hot object from getting blocked by syncs.

### 4.5 Garbage Collection

As discussed in §3.5, masters send garbage collection RPCs for synced updates to their witnesses. Right after syncing to backups, masters send **gc** RPCs (in Figure 4) that indicates the client requests that were just synced.

To identify client requests for removal, CURP uses 64-bit key hashes and RPC IDs assigned by RIFL [10]. Upon receiving a **gc** RPC, a witness locates the sets of slots using the *keyHash*es and resets the slots whose occupying requests have the matching RPC IDs. Witnesses ignore *keyHash*es and *rpcIds* that are not found since the record RPCs might have been rejected. For client requests that mutate multiple objects, **gc** RPCs include multiple $\langle$*keyHash*, *rpcIds*$\rangle$ pairs for all affected objects, so that witnesses can clear all slots occupied by the request.

Although the described garbage collection can clean up most records, some slots may be left uncollected: if a client crashes before sending the update request to the master, or if the **record** RPC is delayed significantly and arrives after the master finished garbage collection for the update. Uncollected garbage will cause witnesses to indefinitely reject requests with the same keys.

Witnesses detect such uncollected records and ask masters to retry garbage collection for them. When it rejects a **record**, a witness recognizes the existing record as uncollected garbage if there have been many garbage collections since the record was written (three is a good number if a master performs only one **gc** RPC at a time). Witnesses notify masters of the requests that are suspected as uncollected garbage through the response messages of **gc** RPCs; then the masters retry the requests (most likely filtered by RIFL), sync to backups, and include them in the next **gc** requests.

### 4.6 Recovery Steps

To recover a crashed master, CURP first restores data from backups and then replays requests from a witness. To fetch the requests to replay, the new master sends a **getRecoveryData** RPC (in Figure 4), which has two effects: (1) it irreversibly sets the witness into recovery mode, so that the data in the witness will never change, (2) it provides the entire list of client requests saved in the witness.

With the provided requests, the new master replays all of them. Since operations already recovered from backups will be filtered out by RIFL [10], the replay step finishes very quickly. In total, CURP increases recovery time by the execution time for a few requests plus 2 RTT (1 RTT for **getRecoveryData** and another RTT for backup sync after replay).

### 4.7 Zombies

For a fault-tolerant system to be consistent, it must neutralize *zombies*. A zombie is a server that has been determined to have crashed, so some other server has taken over its functions, but the server has not actually crashed (e.g., it may have suffered temporary network connectivity problems). Clients may continue to communicate with zombies; reads or updates accepted by a zombie may be inconsistent with the state of the replacement server.

CURP assumes that the underlying system already has mechanisms to neutralize zombies (e.g., by asking backups to reject replication requests from a crashed master [16]). The witness mechanism provides additional safeguards. If a zombie responds to a client request without waiting for replication, then the client must communicate with all witnesses before completing the request. If it succeeds before the witness data has been replayed during recovery, then the update will be reflected of the new master. If the client contacts a witness after its data has been replayed, the witness will reject the request; the client will then discover that the old master has crashed and reissue its request to the new master. Thus, the witness mechanism does not create new safety issues with respect to zombies.

### 4.8 Modifications to RIFL

RIFL is a mechanism for detecting duplicate invocations of RPCs. With RIFL, masters make a durable *completion record* of each RPC that updates state, which includes the RPC result. The completion record survives crashes and can be used to detect duplicate invocations of the RPC. When a duplicate is detected, the master skips the execution of the RPC and returns the result from the completion record.

RIFL has two mechanisms for garbage collecting completion records: (1) on RPC requests, clients piggyback acknowledgments of the results of their previous requests (so servers can safely delete these completion records), and (2) clients maintain leases in a central server; if a client's lease expires, masters can delete all completion records for that client. Both of these must be modified to work with CURP.

Since both garbage collection mechanisms assume that retries always come from the same client that made the original request, RIFL must be modified to accommodate retries from witnesses. Firstly, once clients acknowledge the receipts of results, masters remove their completion records and start to ignore (not returning results) the duplicate requests. Since replays from witnesses happen in random orders, acknowledgements piggybacked on later requests can make masters to ignore the replay of earlier requests. Thus, clients' acknowledgments included in RPC requests must be ignored during recovery from witnesses.

Secondly, if a client crashes and its lease expires, masters remove all of the completion records for the client; then any requests from the expired client are ignored. This can be a problem in CURP since the replay of the expired client's requests will be ignored during witness-based recovery. To prevent this, masters must sync all operations to backups before expiring a client lease. In practice, the period of syncs is much smaller than the grace period between the time of a client crash and the time of its lease expiration; so, most systems are safe automatically.

| | RAMCloud cluster | Redis cluster |
|---|---|---|
| CPU | Xeon X3470 (4x2.93 GHz) | Xeon D-1548 (8x2.0 GHz) |
| RAM | 24 GB DDR3 at 800 MHz | 64 GB DDR4 |
| Flash Disks | 2x Samsung 850 PRO SSDs (256 GB) | Toshiba NVMe flash (512 GB) |
| NIC | Mellanox ConnectX-2 InfiniBand HCA (PCIe 2.0) | Mellanox ConnectX-3 10 Gbps NIC (PCIe 3.0) |
| Switch | Mellanox SX6036 | HPE 45XGc |
| OS | Linux 3.16.0-4-amd64 | Linux 3.13.0-100-generic |

**Table 1:** The server hardware configuration for benchmarks.

# 5 Evaluation

We evaluated CURP by implementing it in the RAMCloud and Redis storage systems, which have very different backup mechanisms. First, using the RAMCloud implementation, we show that CURP improves the performance of consistently replicated systems. Second, with the Redis implementation, we demonstrate that CURP can make strong consistency affordable in a system where it had previously been too expensive for practical use.

## 5.1 RAMCloud Performance Improvements with CURP

RAMCloud [16] is a large-scale low latency distributed key-value store, which primarily focuses on reducing latency. Small read operations take 5 μs, and small writes take 14 μs. By default, RAMCloud replicates each new write to 3 backups, which asynchronously flush data into local drives. Although replicated data are stored in slow disk (for cost saving), RAMCloud features a technique to allow fast recovery from a master crash (it recovers within a few seconds) [15].

With the RAMCloud implementation of CURP, we answered the following questions:

- How does CURP improve RAMCloud's latency and throughput?
- How many resources do witness servers in CURP consume?
- Will CURP be performant under highly-skewed workloads with hot keys?

Our evaluations using the RAMCloud implementation were conducted on a cluster of machines with the specifications shown in Table 1. All measurements used InfiniBand networking and RAMCloud's fastest transport, which bypasses the kernel and communicates directly with InfiniBand NICs. Our CURP implementation kept RAMCloud's fast crash recovery [15], which recovers from master crashes within a few seconds using data stored on backup disks. Servers were configured to replicate data to 1–3 different backups (and 1–3 witnesses for CURP results), indicated as a replication factor $f$. The log cleaner of RAMCloud did not run in any measurements; in a production system, the log cleaner can reduce the throughput.

For RAMCloud, CURP moved backup syncs out of the critical path of write operations. This decoupling not only improved latency but also improved the throughput of RAMCloud writes.

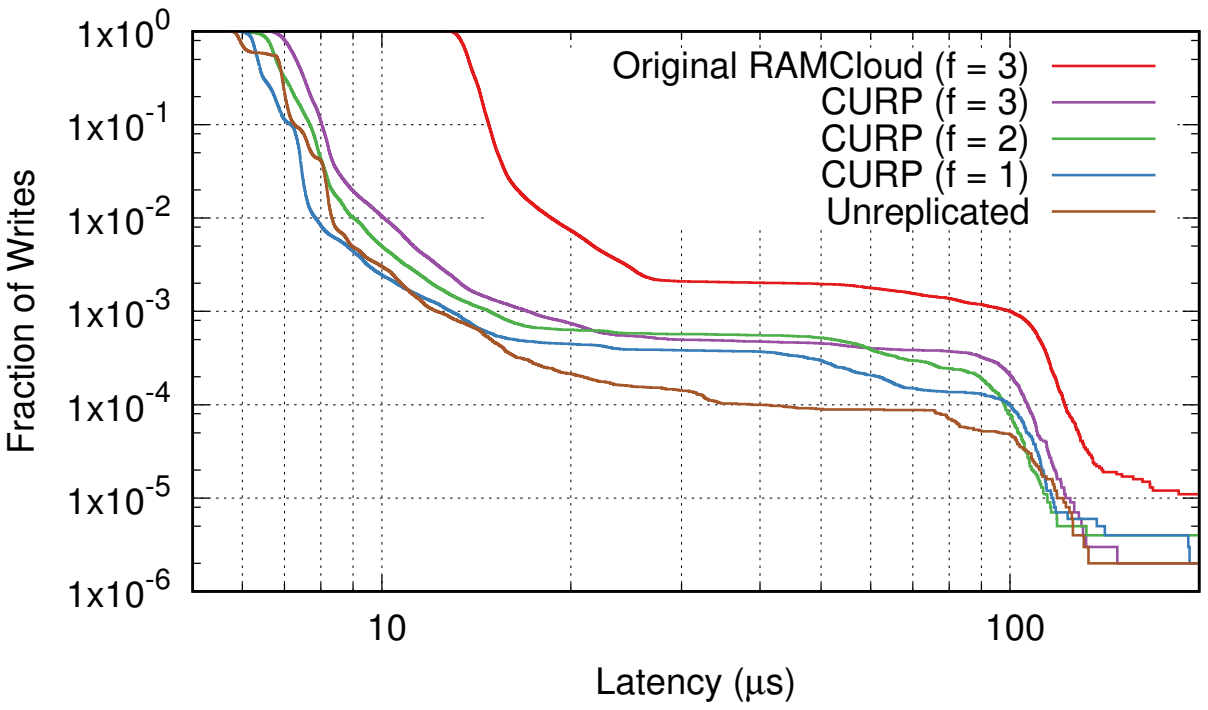

**Figure 5:** Complementary cumulative distribution of latency for 100B random RAMCloud write with CURP. Writes were issued sequentially by a single client to a single server, which batches 50 writes between syncs. A point $(x, y)$ indicates that $y$ of the 1M measured writes took at least $x$ μs to complete. $f$ refers to fault tolerance level (i.e. number of backups and witnesses). "Original RAMCloud" refers to the base RAMCloud system before adopting CURP. "Unreplicated" refers to RAMCloud without any replication. The median latency for synchronous, CURP ($f = 3$) and unreplicated writes were 13.8 μs, 7.3 μs, and 6.9 μs respectively.

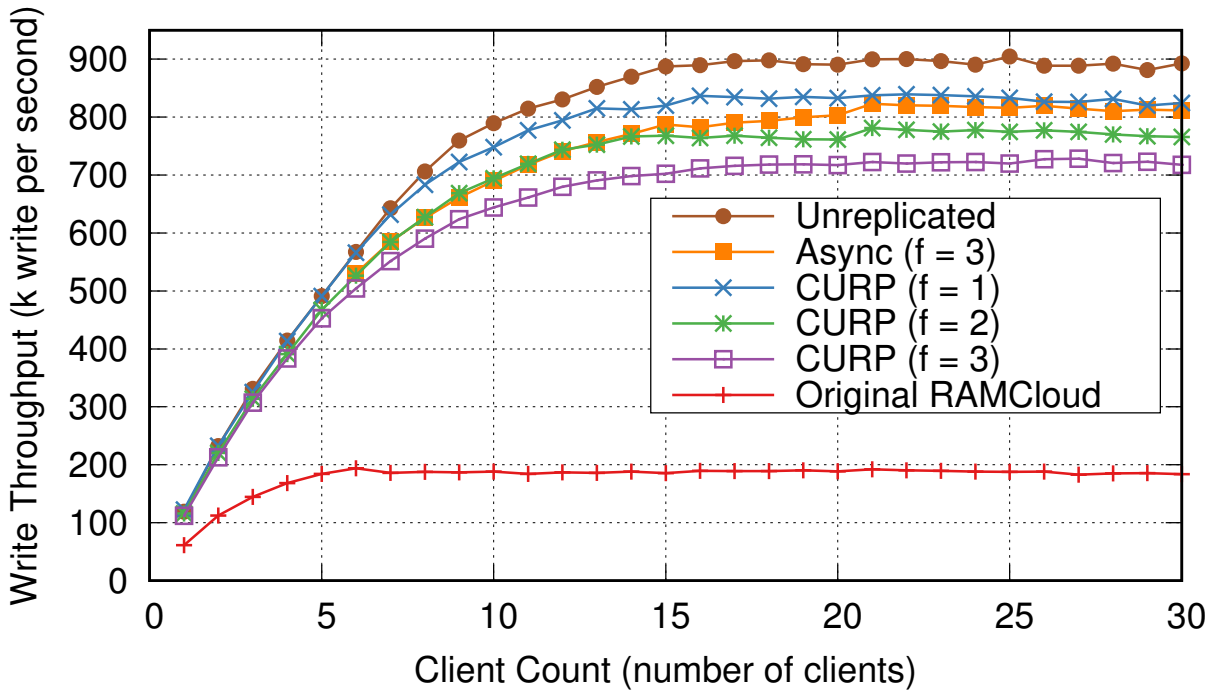

**Figure 6:** The aggregate throughput for one server serving 100B RAMCloud writes with CURP, as a function of the number of clients. Each client repeatedly issued random writes back to back to a single server, which batches 50 writes before syncs. Each experiment was run 15 times, and median values are displayed. "Original RAMCloud" refers to the base RAMCloud system before adding CURP. "Unreplicated" refers to RAMCloud without any replication. In "Async" RAMCloud, masters return to clients before backup syncs, and clients complete writes without replication to witnesses or backups.

Figure 5 shows the latency of RAMCloud write operation before and after applying CURP. CURP cuts the median write latencies in half. Even the tail latencies are improved overall. When compared to unreplicated RAMCloud, CURP with one or two replicas did not add noticeable latency overhead, and CURP with three replicas adds 0.4 μs to median latency. This is because witnesses process record RPCs much faster than masters process update RPCs, so clients can send and receive record RPCs while waiting for responses for write RPCs.

Figure 6 shows the single server throughput of write operations before CURP and after CURP by varying the number of clients. The server batches 50 writes before starting a sync. By batching backup syncs, CURP improves throughput by about 4x. When compared to unreplicated RAMCloud, adding an additional CURP replica drops throughput by ∼6%.

To illustrate the overhead of CURP on throughput (e.g. sending gc RPCs to witnesses), we measured RAMCloud with asynchronous replication, which is identical to CURP except that it does not record information on witnesses. Achieving strong consistency with CURP reduces throughput by 10% for 3-way replication. In all configurations except the original RAMCloud, masters are bottlenecked by a dispatch thread (RAMCloud masters dedicate a thread to receive client requests from the NIC, dispatch the requests to worker threads, and send replies to clients). Thus, witness gc RPCs burden the already bottlenecked dispatch thread; this is the main reason why throughput drops under CURP.

### 5.2 Resource Consumption by Witness Servers

Each witness server implemented in RAMCloud can handle 1270k record requests per second with occasional garbage collection requests (1 every 50 writes) from master servers. This witness server runs on a single thread and consumes 1 hyper-thread core at max throughput. Considering that each RAMCloud master server uses 8 hyper-thread cores to achieve 728k writes per second, adding 1 witness increases the total CPU resources consumed by RAMCloud by 7%. However, CURP reduces the number of distinct backup operations performed by masters, because it enables batching; this offsets most of the cost of the witness requests (both backup and witness operations are so simple that most of their cost is the fixed cost of handling an RPC; a batched replication request costs about the same as a simple one).

The second resource overhead is memory usage. Each witness server allocates 4096 request storage slots for each associated master, and each storage slot is 2KB. With additional metadata, the total memory overhead per master-witness pair is around 9MB.

The third issue is network traffic amplification. In CURP, each client request is replicated twice to witnesses and backups. With 3-way replication, CURP increases network bandwidth use by 75% (in the original RAMCloud, a client request is transferred over network to a master and 3 backups).

### 5.3 Impact of Highly-Skewed Workload with Hot Keys

CURP cannot complete an operation in 1 RTT unless it is commutative with other unsynced ones. In NoSQL systems, it can only allow 1 unsynced update per key for each batched backup sync. This may impact performance for workloads with popular keys that are frequently updated. To measure the impact of such hot keys, we measured RAMCloud write latencies with CURP using a highly-skewed Zipfian distribution with 1M objects and a parameter of 0.99 (these are the defaults for the YCSB-A and YCSB-B workloads [3]).

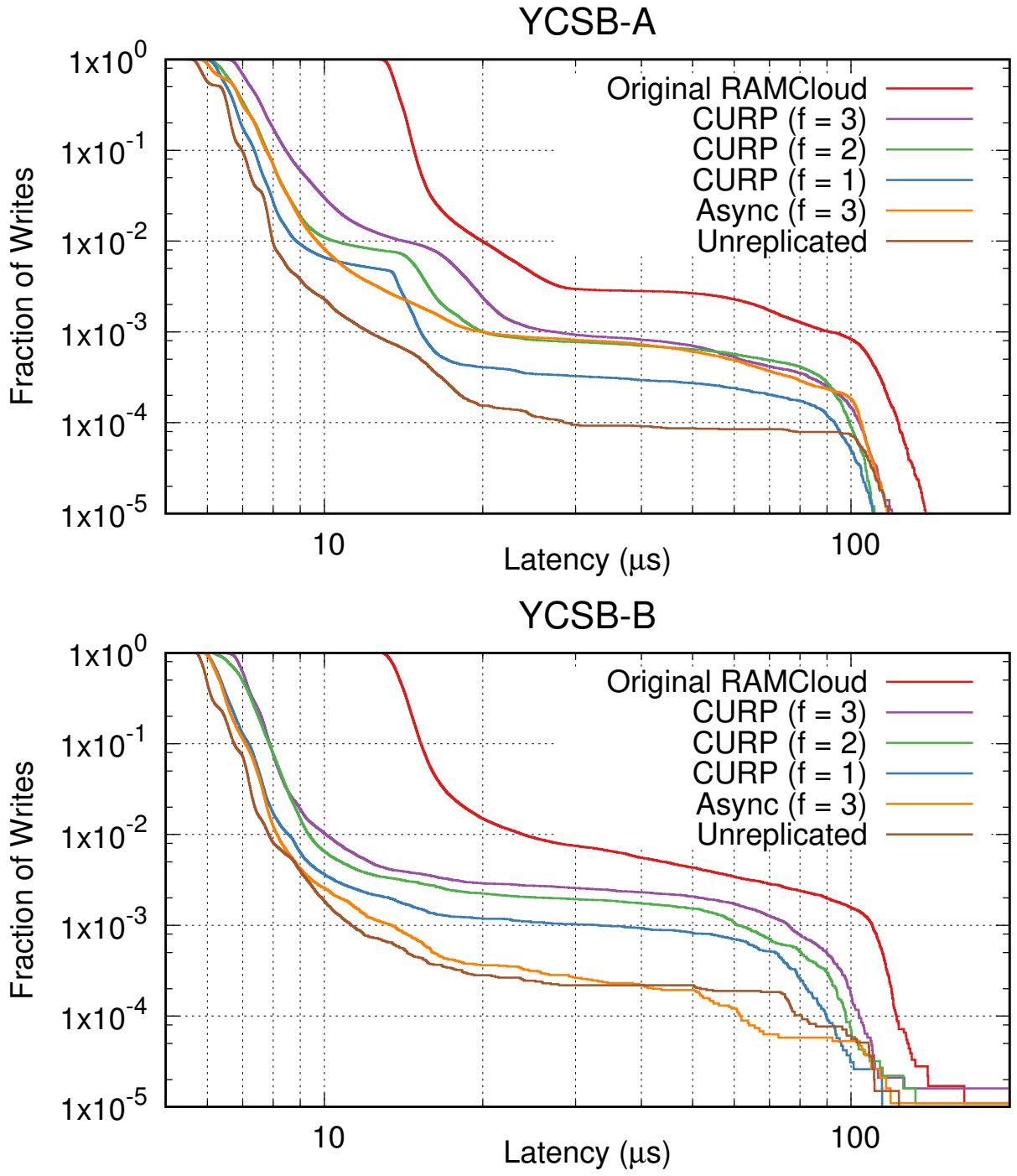

**Figure 7:** Complementary cumulative distribution of latency for YCSB-A and YCSB-B workloads with CURP. A single client issued read or write operations to a single server, which batches 50 writes before syncs. A point $(x,y)$ indicates that $y$ of measured writes took at least $x$ µs to complete.

Figure 7 shows RAMCloud write latencies with and without CURP under the YCSB-A and YCSB-B workloads. CURP maintains low latencies even with the highly-skewed workloads. Even when writes are conflicting (∼1%), latencies stay at 2 RTTs (lines in Figure 7 kink at around 14 µs); in most cases when witnesses reject record RPCs, the masters also detect the conflicts and sync before returning to clients (operations take 2 RTTs total), so clients do not have to send sync RPCs to the masters (which could incur an extra 1 or 2 RTT).

Since CURP delays syncs for batching, the highly-skewed Zipfian distribution used by YCSB made key collisions more likely. To minimize this contention and achieve 1 RTT, masters can be set to start syncing immediately after responding back to clients, which will reduce thoughput ∼20% (See §C.1 for details).

### 5.4 Using CURP to Make Redis Consistent and Durabile

Redis [19] is another low-latency in-memory key-value store, where values are data structures, such as lists, sets, etc. For Redis, the only way to achieve durability and consistency after crashes is to log client requests to an append-only file and invoke fsync before responding to clients. However, fsyncs can take several milliseconds, which is a 10–100x performance penalty. As a result, most Redis applications do not use synchronous mode; they use Redis as a cache with no durability guarantees. Redis also offers replication to multiple servers, but the replication mechanism is asynchronous, so updates can be lost after crashes; as a result, this feature is not widely used either.

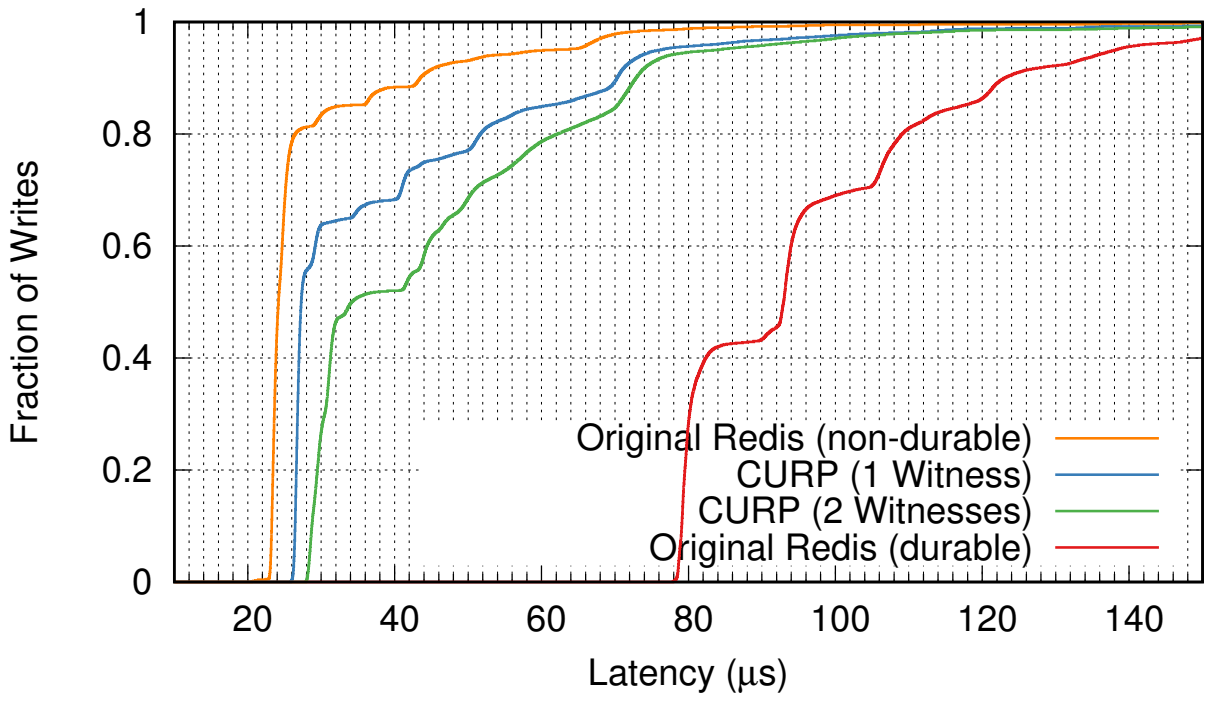

**Figure 8:** Cumulative distribution of latency for 100B random Redis SET requests with CURP. Writes were issued sequentially by a single client to a single Redis server. CURP used one or two additional Redis servers as witnesses. "Original Redis (durable)" refers to the base Redis without CURP, configured to invoke fsync on a backup file before replying to clients.

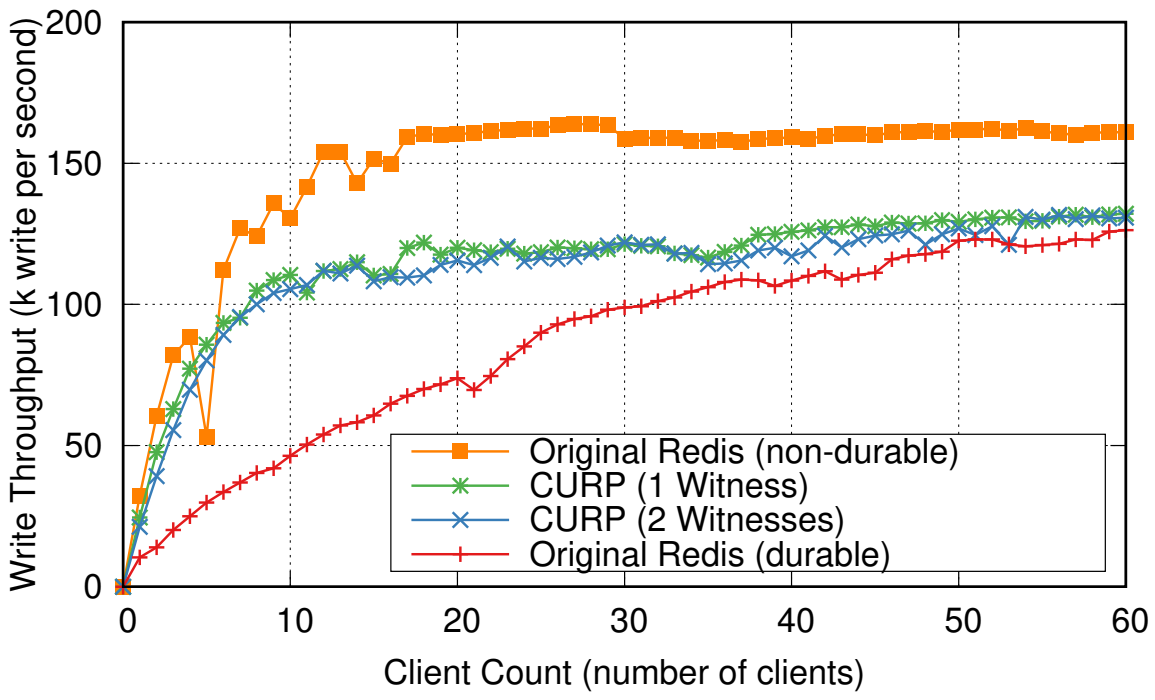

**Figure 9:** The aggregate throughput for one server serving 100B Redis SET operations with CURP, as a function of the number of clients. Each client repeatedly issued random writes back to back to a single server. "Original Redis (durable)" refers to the base Redis without CURP, but configured to invoke fsync before replying to clients. Original Redis processes requests from multiple clients, fsyncs once per eventloop, and replies to all clients.

For this experiment, we used CURP to hide the cost of Redis' logging mechanism: we modified Redis to record operations on witnesses, so that operations can return without waiting for log syncs. Log data is then written asynchronously in the background. The result is a system with durability and consistency, but with performance equivalent to a system lacking both of these properties. In this experiment the log data is not replicated, but the same mechanism could be used to replicate the log data as well.

With the Redis implementation of CURP, we answered the following questions:

- Can CURP transform a fast in-memory cache into a strongly-consistent durable storage system without degrading performance?
- How wide a range of operations can CURP support?

Measurements of the Redis implementation were conducted on a cluster of machines in CloudLab [18], whose specifications are in Table 1. All measurements were collected using 10 Gbps networking and NVMe SSDs for Redis backup files. Linux fsync on the NVMe SSDs takes around 50$\sim$100 µs; systems with SATA3 SSDs will perform worse with the fsync-always option. Automatic re-writing of the append-only file is turned off for all Redis measurements.

For the Redis implementation, we used Redis 3.2.8 for servers and "C++ Client" [22] for clients. We modified "C++ Client" to construct Redis requests more quickly.

Figure 8 shows the performance of Redis before and after adding CURP to its local logging mechanism; it graphs the cumulative distribution of latencies for Redis SET operations. After applying CURP (using 1 witness server), the median latency increased by 3 µs, which is 12%. The additional cost is caused primarily by the extra syscalls for send and recv on the TCP socket used to communicate with the witness; each syscall took around 2.5 µs.

When a second witness server is added in Figure 8, latency increases significantly. This occurs because the Redis RPC system has relatively high tail latency. Even for the non-durable original Redis system, which makes only a single RPC request per operation, latency degrades rapidly above the 80th percentile. With two witnesses, CURP must wait for three RPCs to finish (the original to the server, plus two witness RPCs). At least one of these is likely to experience high tail latency and slow down the overall completion. We didn't see a similar effect in RAMCloud because its latency is consistent out to the 99th percentile: when issuing three concurrent RPCs, it is unlikely that any of them will experience high latency.

Figure 9 shows the throughput of Redis SET operations for a single Redis server with varying numbers of clients. Applying CURP reduced the throughput of Redis about 18%. With a large number of clients, the original synchronous form of Redis can offer throughput approaching non-durable Redis. The reason for this is that Redis batches fsyncs in synchronous made: in each cycle through its event loop, it processes all of the requests waiting on its incoming sockets, issues a single fsync, then responds to all of those requests. The disadvantage of this approach is that it results in very high latency for clients; see §C.2 for details.

### 5.5 Applicability of CURP

CURP can be applied to a variety of operations, not just write operations in key-value stores. Redis supports many data structures, such as strings, hashmaps, lists, counters, and so on. All of these update operations can benefit from CURP. Since each data structure is assigned to a specific key, CURP can execute many update operations on different keys without blocking on syncs.

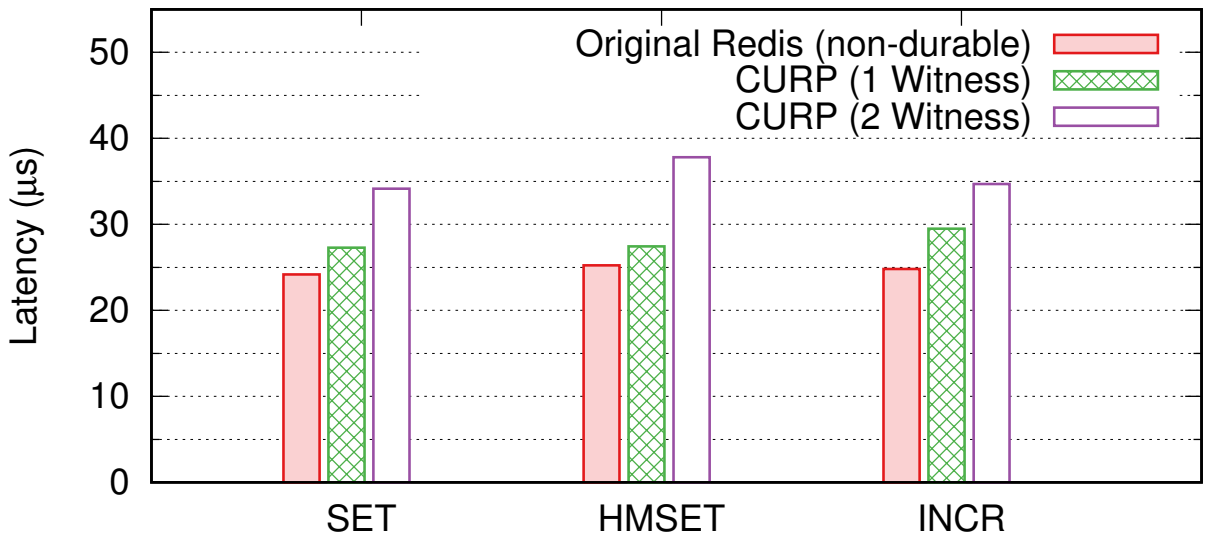

**Figure 10:** Median latencies before and after applying CURP on various Redis commands. All experiments select a random 30B key over 2M unique keys. SET used 100B random values, and each HMSET operation sets 1 member with a 100B value. The member key was 1B. Commands were issued sequentially by a single client to a single Redis server, with one or two additional Redis witness servers in CURP.

Figure 10 shows the median latency with and without CURP on three different Redis commands: SET, where ASCII data is written to a string data structure; HMSET, where data is written to a member of a hashmap; and INCR, where an integer counter is incremented. For all three operations, latency overheads were small for CURP with 1 witness. CURP with 2 witnesses increased latency about 10 µs because of tail latency issues. We believe that the TCP transport library used by the C++ client is inefficient for waiting for multiple responses concurrently, and we will continue to investigate this.

## 6 Related work

*Fast Paxos* [9] allows clients to complete operations in 1.5 RTTs; clients can send their operations directly to replicas. If the operations are accepted, replicas send acknowledgments to a strong leader, and then the leader executes the operations. However, if multiple clients concurrently send operations, they may contend on a command slot, resulting in extra communications to resolve the conflict.

*Generalized Paxos* [8] uses commutativity to resolve the contention problem of Fast Paxos. To reduce contention for command slots, Generalized Paxos groups commutative requests from concurrent clients into an unordered set, and it only orders between sets. This is very similar to CURP; CURP requires unsynced operations to be commutative. However, the leader in Generalized Paxos cannot execute an operation until it has heard from the replicas, so Generalized Paxos requires 1.5 RTTs, as opposed to 1 RTT for CURP.

*Egalitarian Paxos* (EPaxos) [12] relies on commutativity to allow multiple leaders to propose and execute operations concurrently. This approach improved throughput. In geo-replicated environments, EPaxos allows clients to choose a nearby replica as leader, so operations can complete in 1 wide-area RTT. However, EPaxos's fungible leaders complicate read operations. To avoid replicating read requests, consensus protocols with strong leaders usually use a lease for the leader; a leader with a valid lease can execute read operations without replication. Since EPaxos does not have strong leaders, they have to use full consensus for read operations as well, or use fine granular leases per object while sacrificing the flexibility of choosing a nearby leader. On the other hand, CURP can just execute read operations in masters or even in backups with help of witnesses (see §A.1 for how it works). Also, without dealing with the complexity of multiple leaders, CURP removes most of the replication overhead in masters by batching syncs to backups.

*Speculative Paxos* [17] and *Network-Ordered Paxos* (NOPaxos) [11] reduced latency to 1 RTT by serializing client requests through network devices. In Speculative Paxos, client request packets detour through a root-layer network switch to mostly order them. NOPaxos routes client request packets through a network processor called a sequencer (implemented by root layer switches or middleboxes). In both protocols, the network must support SDN to make requests to detour through a single special network device; so, they can be deployed only in specialized environments (e.g. privately-owned datacenters). Also, due to detouring of packets, they actually add latency overhead over unreplicated systems; Speculative Paxos (∼25 µs) or NOPaxos(∼16 µs) have higher latency overhead compared to CURP (within 0.4 µs).

## 7 Conclusion

One of the most effective ways to improve the performance of computer systems is by identifying operations that can be performed concurrently. In this paper we have uncovered an opportunity for introducing concurrency into mechanisms for consistent replication. By exploiting the commutativity of operations, replication without ordering can be performed in parallel with sending requests to an execution server. This general approach can be applied to improve a variety of replication mechanisms, including primary-backup approaches and consensus protocols with strong leaders. We presented Consistent Unordered Replication Protocol (CURP), which supplements standard primary-backup replication mechanisms. CURP reduces the latency to complete operations from 2 RTTs to 1 RTT while retaining strong consistency. We implemented CURP in RAMCloud and Redis to demonstrate its benefits.

# A Extra Discussions

## A.1 Consistent Reads from Backups

Servicing read operations from backups is benefitial since it reduces the load in masters and can provide better latency in a geo-replicated environment (clients can read from a backup in the same region, providing 0 wide-area RTT). However, naiively reading from backups can violate linerizability since update operations in CURP can complete before syncing to backups.

To avoid reading the stale value and risking linearizability, clients in CURP can use a nearby witness to check whether the value read from a nearby backup is up to date. To perform a consistent read, a client must ask a witness whether the read operation commutes with the operations currently saved in the witness. If it commutes, the client is assured that the value read from the backup will be up to date. If it doesn't commute (i.e. the witness retains a write request on the key being read), the value read from the backup might be stale. In this case, the client must read from the master.

This technique is safe because clients can complete an operation only if it is synced to *all* backups or recorded in *all* witnesses. Thus, for every completed operation, either the backup has the up-to-date value or the witness has a record of the operation. By checking whether a witness has a noncommutative request, the reader client can tell whether there is an update operation which is completed but not yet flushed to the backup. If there is an ongoing update, the client must read from the master; if not, the client is assured that the value read from the backup is a consistent value.

Additionally, while an update is being synced to backups, other clients in system cannot read the new unsynced value. As discussed in §3.2.3, masters keep unsynced operations commutative, so the read operations on the unsynced new value (which are not commutative with the update operation) cause the master to sync the new value to all backups before returning to the clients. Thus, any reads on the new value from the master cannot complete while backups have the stale value.

Therefore, between the time when the operation completes (which is after it is written in witnesses and executed in the master) and the time when it is synced to backups, no clients can read either the old or new value; before the completion of the operation, clients can read only the old value; after it is synced to backups, clients can read only the new value.

## A.2 Extending CURP to Consensus Protocols

This section illustrates how CURP can be extended to reduce the latency of consensus protocols. CURP can be integrated in most consensus protocols with strong leaders (e.g. Raft [14], Viewstamped Replication [13]). In such protocols, clients send requests to the current leader replica, which serializes the requests into its command log. The leader then replicates its command log to a majority of replicas before executing the requests and replying back to clients with the results. This process takes 2 RTTs, and CURP can reduce it to 1 RTT.

As in primary-backup replication, CURP on consensus allows clients to replicate requests to witnesses in parallel with sending requests to the leader; the leader then speculatively executes the requests and responds to clients before replicating the requests to a quorum of replicas. A client can complete an operation if it is accepted by a superquorum of witnesses or committed in a quorum of replicas.

To mask $f$ failures, consensus protocols use $2f+1$ replicas, and systems stay available with $f$ failed replicas. For the same guarantee, CURP also uses $2f+1$ replicas, but each replica also has a witness component in addition to existing components for consensus. Although CURP can proceed with $f+1$ available replicas, it needs $f+\lceil f/2\rceil+1$ replicas (for superquorum of witnesses) to use 1 RTT operations. With less than $f+\lceil f/2\rceil+1$ replicas, clients must ask masters to commit operations in $f+1$ replicas before returning result (2 RTTs).

Like masters in regular CURP, leader replicas execute operations speculatively if they are commutative with existing unsynced operations; for an incoming client request, a leader serializes it into the command log, executes it, and responds to the client before committing it in a majority of replicas.

For clients to complete an operation in 1 RTT, it must be recorded in a *superquorum* of $f+\lceil f/2\rceil+1$ witnesses. The reason why CURP needs a superquorum instead of a simple majority is to ensure commutativity of replays from witnesses during recovery. During recovery, only $f+1$ out of $2f+1$ replicas (each of which embeds a witness) might be available. If a client could complete an operation after recording to $f+1$ witnesses, the completed operation may exist in only 1 witness out of available $f+1$ witnesses during recovery (since intersection of two quorum is 1 replica). If the other $f$ witnesses accepted other operations that are not commutative with the completed operation (since each witness enforce commutativity individually), recovery cannot distinguish which one is the completed one; executing all appearing in any $f+1$ witnesses is also not safe since they are not commutative, so they must be replayed in a correct order.

For correctness, the client requests replayed from witnesses during recovery must be *commutative* and *inclusive* of all completed operations that are not yet committed in a majority of replicas. By recording to a superquorum, all completed operations (but not yet committed) are guaranteed to exist in a majority ($\lceil f/2\rceil+1$) of any quorum of $f+1$ witnesses, and any operations that doesn't commute with the completed operations cannot exist in more than $\lfloor f/2 \rfloor$ (less than majority of any quorum). Thus, during recovery, all requests that appear more than a majority ($\lceil f/2 \rceil + 1$) from any quorum of $f+1$ witnesses are guaranteed to be commutative and include all completed operations; so, recovery can replay requests that appear in more than $\lceil f/2 \rceil + 1$ witnesses out of any $f+1$ witnesses.

When leadership changes (e.g. leader election in Raft [14] or view change in Viewstamped Replication [13]), the new leader must recover from witnesses before accepting new operations. To do so, the new leader must collect saved requests in at least $f+1$ witnesses. This collection can be included in the existing data collection (e.g. Raft votes) that is required by most leadership change protocols. As mentioned in previous paragraph, the new leader should only replay client requests that are recorded in more than $\lceil f/2 \rceil + 1$ witnesses to ensure commutativity.

After leadership changes, the state machine of the old leader could have diverged from other replicas due to speculatively executed operations that were not recovered from witnesses. To fix this, the old leader must reload from a checkpoint that does not have speculative executions. However, we can avoid reloading from checkpoints if the leadership change was not because of a crash or disconnect of the old leader; instead of requring old leader to reload from a checkpoint, we can require the new leader to fetch and commit all uncommitted operations in the old leader's command log.

The last problem introduced by speculative execution is clients may use old zombie leaders (which believe they are current leaders). Zombie leaders were not impossible before CURP since an operation must be committed in a majority before being executed and at least one replica would reject the operation. To prevent clients from completing operations with an old (possibly disconnected) leader, they tag record RPCs with a term number (e.g. a Raft term or a view-number in Viewstamped Replication), which increments every time when leadership changes. A witness checks the term number against the term used by its replica (recall that a witness is a part of a consensus replica); if the record RPC has an old term number, the witness rejects the request and tells the client to fetch new leader information.

### A.3 Does CURP Slow Down Reads?

Read operations in CURP may have to wait for the backup sync of the value being read. In the worst-case, one update can block multiple concurrent reads on the same object. We divide reads into two categories and discuss how to mitigate this problem for each type.

The first type of reads is reading values in preparation for subsequent updates, such as conditional writes and transactions. The updates check to ensure that the previously read values have not changed, and the updates abort if any value has changed. Applications of such conditional updates handle aborts by retrying.

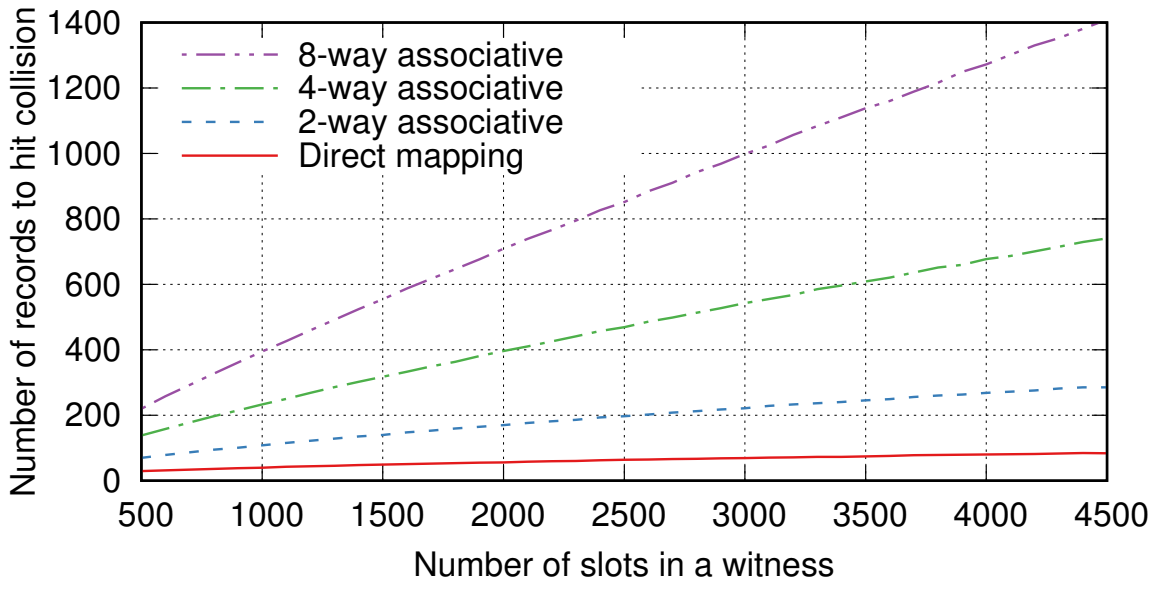

**Figure 11:** Simulation results for the expected number of recordings before a collision occurs in a witness' cache, assuming a random distribution of keys. Each data point is the average of 10000 simulations. Introducing associativity reduces the chance of collisions significantly.

In applications like these, it is safe to allow the initial reads to complete without waiting for durability of the values. The later reads peformed during transaction commit must still wait for durability.

The second type is purely reading with no subsequent updates. In such cases, applications can tolerate slightly stale values. Maintaining multiple versions of values, which is widely used in transaction concurrency control, can resolve the issue of blocking reads. For systems without native multi-version support, a simple cache for keeping the latest durable values can be used to service reads. This cache can be simple and small. In fact, the structure of the durable value cache is same as that of witnesses since the cache only keeps old values for updates completed asynchronously by saving RPC requests in the witness buffer.

## B Implementation Details

### B.1 Why Use Set-associative Cache for Witnesses?

We initially used a direct-mapped cache instead of set-associative cache, but this resulted in a high rate of rejections because of conflicts (i.e. no slot is available for the mapped set). Figure 11 shows the expected number of recordings before a conflict occurs on a witness slot. Using a direct mapping and 4096 total slots, it is expected to have a false conflict after about 80 insertions. Thus, we switched to 4-way associative cache, to reduce witness rejections. We didn't need 8-way associativity (a bit slower than 4-way) since the number of requests in witnesses is already limited by commutativity. (Once a master hits a non-commutative operation and syncs to backups, all saved requests in the witness are garbage collected.)

## C Additional Evaluations

### C.1 RAMCloud's Throughput by Batch Size

Figure 12 shows the single-server throughput of write operations with CURP while varying the aggressiveness of syncs. After introducing CURP, RAMCloud can delay the sync to backups after responding back to clients; delaying and batching sync to backups makes the server more efficient and improves throughput about 4 times. Since RAMCloud allows only one outstanding sync, syncs are naturally batched for around 15 writes even at 1 minimum batch size.

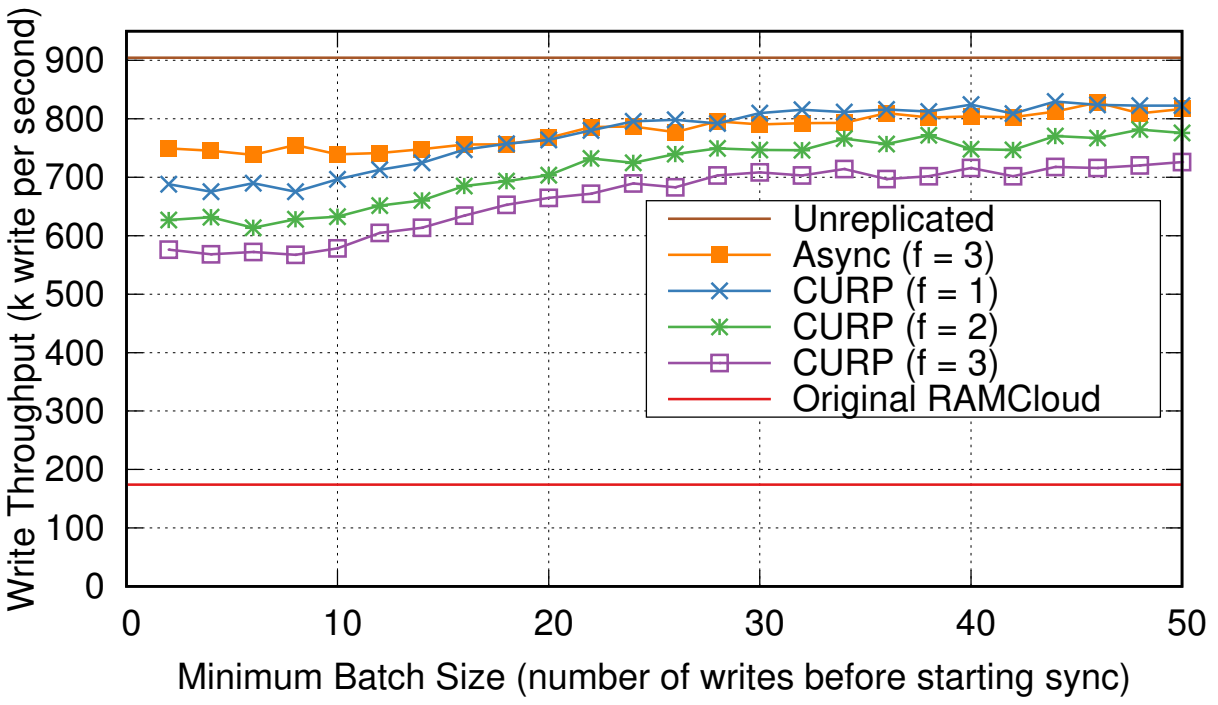

**Figure 12:** The aggregate throughput for one server serving 100B RAMCloud writes with CURP, as a function of sync batch size. Each client repeatedly issued random writes back to back to a single server. "Original RAMCloud" refers to the base RAMCloud system before adding CURP. "Unreplicated" refers to RAMCloud without any replication. Each datapoint was measured 15 times, and median values are displayed.

### C.2 Redis Latency vs. Throughput

Figure 13 shows observed latency during throughput benchmark. Both CURP and non-durable Redis maintains latency low until it reaches 80% of max throughput. The latency of durable Redis increases almost linearly due to bathcing. The original Redis is designed to provide maximum throughput under high load and natively batches fsyncs; for each event-loop cycle, Redis iterates through TCP sockets for all clients and executes all requests from them; after the iteration, Redis fsyncs once and responds to the clients. This batching amortizes the cost of fsync, and throughput of durable Redis approaches that of non-durable Redis as the number of clients increases. However, this batching adds extra delay before responding back to clients, so latency increases up linearly.

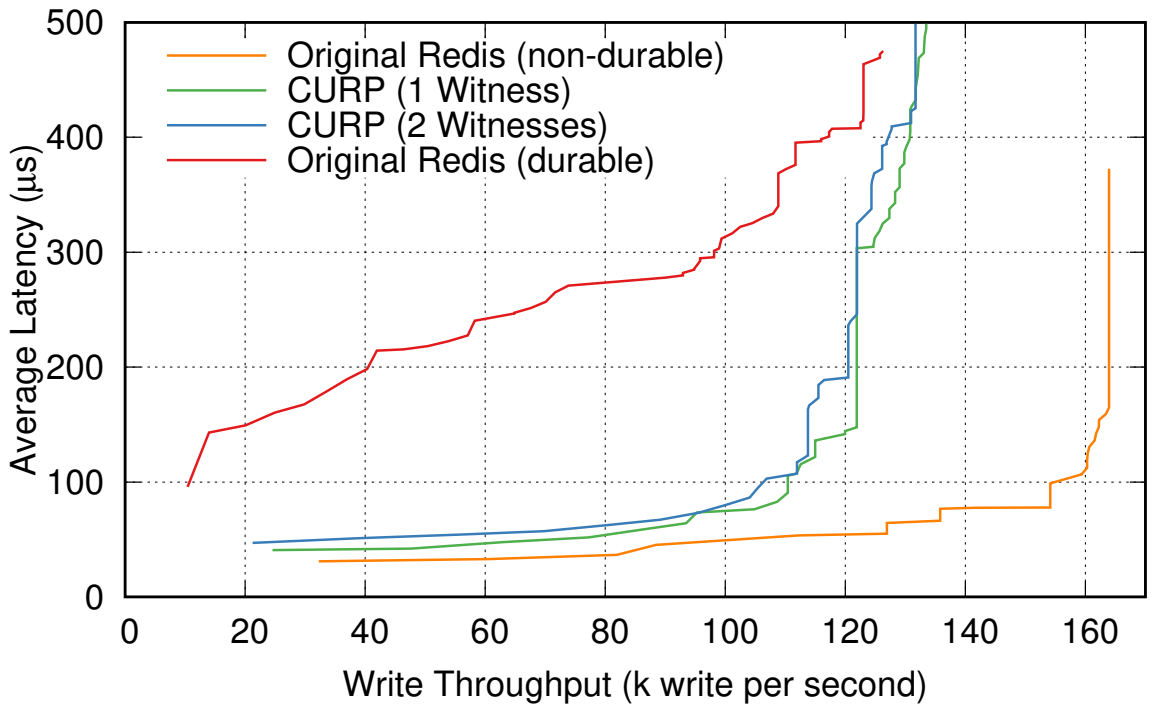

**Figure 13:** Observed latency at a specific throughput level for one server serving 100B Redis SET operations with CURP. "Original Redis (durable)" refers to the base Redis without CURP, but configured to invoke fsync before replying to clients. Original Redis processes requests from multiple clients, fsyncs once per eventloop, and replies to all clients.